# Spectral Properties of Fermi Blazars and their Unification Schemes

E.U. Iyida[*], F.C. Odo, A.E. Chukwude and A.A. Ubachukwu

Astronomy and Astrophysics Research Group, Department of Physics and Astronomy, University of Nigeria, Nsukka,
Department of Physics and Astronomy, Faculty of Physical sciences, University of Nigeria, Nsukka, Nigeria
*email: uzochukwu.iyida@unn.edu.ng

**Abstract**

We use the distributions of spectral indices ($\alpha_v$) of a large homogenous sample of *Fermi*-detected blazars to re-investigate the relationship between flat spectrum radio quasars (FSRQs) and subclasses of BL Lac objects (BL Lacs). We compute the broadband synchrotron and Compton spectral indices from radio-to-X-ray and X-ray to γ-ray bands, respectively. Analyses of our data show continuity in the distributions of the spectral indices from FSRQs to HSP through LSP and ISP subclasses of BL Lacs. We find from γ-ray luminosity distribution that the jetted radio galaxies form the low-luminosity tail of the distribution, suggestive that the sequence can be extended to the young jetted galaxy populations. We observe a significant difference in the shape of Compton and synchrotron spectra: significant anti-correlation ($r \sim -0.80$) exists between the broadband Compton and synchrotron spectral indices. Furthermore, the broadband spectral indices vary significantly with redshift ($z$) at low redshift ($z < 0.3$) and remain fairly constant at high ($z \geq 0.3$) redshift. The trend of the variations suggests a form of evolutionary connection between subclasses of blazars. Thus, while selection effect may be significant at low redshift, evolutionary sequence can also be important. Our results are not only consistent with a unified scheme for blazars and their young jetted galaxy counterparts but also suggest that the broadband spectral sequence of blazars is not a secondary effect of redshift dependence.

## 1. Introduction

Blazars are exciting subclass of active galactic nuclei (AGNs) that are thought to be very bright and inconsistent sources of high energy (Abdo *et al.* 2009, 2010a; Yefimov, 2011). They reportedly have other outstanding observational features like large and variable luminosity, continuous spectral energy distributions (SEDs), apparent superluminal motions, high and variable polarization and very active GeV to TeV γ-ray emissions (Andruchow *et al.* 2005; Abdo *et al.* 2010b; Yang *et al.* 2018). The strong non-thermal continuum emissions from these objects cover the whole electromagnetic spectrum from a relativistic jet observed along the line of sight, thus, giving rise to relativistic magnification (Urry and Padovani 1995; Blandford and Rees 1978). Blazars are derived from two distinct AGN populations, namely: flat spectrum radio quasars (FSRQs) and BL Lacertae objects (BL Lacs). The major difference between the two broad subclasses depends on the nature of their optical emissions. While FSRQs are known to have strong emission lines, similar to those of normal quasars, BL Lacs are characterized by very weak or no emission lines, with typical equivalent width (EW) < 5 Å (see e.g. Scarpa and Falomo 1997; Padovani, 2007). According to the unified schemes, blazars are simply radio galaxies with their radio jets forming a small angle with respect to the line of sight (Urry and Padovani 1995). The long-standing orientation-based unification scheme for radio-loud AGNs proposes BL Lacs and FSRQs as counterparts to Fanaroff-Riley type I and II (FR I and FR II) radio galaxies, respectively, based on similar spectra, morphologies and range in extended radio luminosity (Urry and Padovani 1995). However, violations to this scheme are well-known and findings of powerful FR II-like BL Lacs, low radio-power FSRQs and BL Lacs exhibiting broad lines in low continuum states (e.g. Kharb *et al.*2010; Raiteri *et al.*2016) appear to break this simple dichotomy between the powerful edge-brightened FR II galaxies and low-power edge-darkened FR Is.

The typical spectral energy distribution (SED) of blazars exhibits double peaks, usually attributed to synchrotron and inverse Compton radiation from the relativistic jets, which dominate in blazars due to the alignment of the jet axis close to the line of sight (Böttcher 2007). For majority of blazars, the synchrotron peaks are usually located between the radio to X-ray frequencies (Giommi et al. 1995) while the Compton peaks are located in the γ-ray regime (e.g. Finke 2013). However, some blazars are known to have Compton peaks in TeV γ-ray frequencies (Li *et al.* 2010). Obviously, the synchrotron peak frequency ($v_{peak}$) may take on a wide range of values (from $10^{12}$ to $10^{20}$ Hz at the extremes) and is one of the principal ways to classify individual blazars. While the FSRQs are found to have low $v_{peak}$ ($<10^{14.5}$ Hz), BL Lacs are found to span the entire range (e.g. Ackerman *et al.* 2011).Thus, BL Lacs are conveniently classified into high-synchrotron peaked (HSP), with synchrotron peak frequency in the range $\log v_{peak}^{syn}$ (Hz) > 15, intermediate synchrotron peaked (ISP) with $14 \leq \log v_{peak}^{syn}$ (Hz) $\leq 15$ and low synchrotron peaked (LSP) with $\log v_{peak}^{syn}$ (Hz) < 14, (e.g. Abdo et al. 2010a; Ackermann *et al.* 2011).

The mechanisms of the continuum emission as well as the relationship between the different subclasses of the blazars have been the subject of several investigations and a wide range of results have been reported (e.g. Cosmastri *et al.,* 1997; Fossati *et al.,* 1998; Xie *et al.,* 2001; Cellone *et al.,* 2007; Abdo *et al.* 2010c; Gaur *et al.,* 2010; Lister *et al.,* 2011; Urry 2011). However, blazar phenomena are often interpreted in the framework of relativistic beaming (Kollagaard, 1994) of radiation from the jets pointing close to the line of sight, which suggests an orientation connection between different classes of the blazars. On the other hand, it is widely argued that FSRQs and BL Lacs are different manifestations of the same physical process that differ only by bolometric luminosity (e.g. Fossati *et al.* 1998; Ghisellini *et al.* 1998). Recent results (e.g. Odo *et al.* 2017) also seem to provide substantial evidence that FSRQs can evolve into BL Lacs through luminosity evolution. In fact, using an analytic model for the synchrotron-self-Compton and Compton-scattered external radiation from blazar jets, an evolutionary scenario that links these blazar subclasses in terms of a reduction of the black-hole accretion power with time has also been proposed (e.g.Ajello *et al.* 2014). All these suggest a trend in variation of spectral properties of blazars from FSRQs through LSPs and ISPs to the HSPs at the extreme of the continuum (e.g. Fossati *et al.* 1998; Giommi *et al.* 2012).

The spectral energy distributions of these objects display notable stability and follow the unified scheme which argues that the blazar sequence is valid, although the main mechanisms responsible for the relationships among the subclasses remain inconsistent as their individual emission continuum forms are significantly different (Sambruna *et al.,* 1996; Ghisellini *et al.,* 1998). Although the paradigm of synchrotron and inverse Compton scattering by leptonic processes in the jets pointing close to the line of sight has been remarkably successful in modelling the broadband spectral energy distributions of blazars, many fundamental issues remain, including the role of hadronic processes. Moreover, a class of γ-ray-emitting radio galaxies, which are thought to be the misaligned counterparts of blazars, has emerged from the results of the Fermi-Large Area, as well as Cherenkov telescopes (Teng *et al.* 2011; Rani 2019). Soft γ-ray emission has also been detected from a few nearby Seyfert galaxies, though it is not clear whether those γ-rays originate from the nucleus (Teng et al. 2011). Thus, blazars and these misaligned counterparts are believed to make up the MeV – GeV - TeV extragalactic γ-ray background, which are recently suspected of being the sources of high energy cosmic rays (e.g. Palladino *et al.* 2019). These misaligned jetted AGNs are thus expected to be part of the unified sequence for γ-ray loud populations.

An important aspect of the blazar sequence that has gained attention of authors (e.g. Finke, 2013; Giommi *et al.*2013; Nalewajko and Gupta, 2017) is the relationship between low energy and high-energy components of the SED. Fossati et al. (1998) introduced a broad band parameter, namely, γ-ray dominance ($D_g$), defined as the ratio of γ-ray luminosity ($L_g$) to the luminosity at synchrotron peak ($L_{pk}$) and found strong anti-correlation between the parameter and synchrotron peak frequency ($v_{pk}$) for the EGRET blazars, which the authors used to argue for a blazar sequence (see also Odo and Aroh, 2020). Another parameter that relates the low- to high-energy components of the SED is the spectral index (α). This parameter is often defined between two frequencies in the low and high energy regimes (e.g. Abdo et al. 2010a). A major argument against this parameter is that it is somewhat redshift-dependent (e.g. Athreya and Kapahi, 1998), which needs to be corrected for in source samples. Comastri et al. (1997) discovered that there is a significant anti-correlation between the broadband spectral indices from radio to optical ($α_{ro}$) and optical to X-ray ($α_{ox}$) of BL Lac objects and FSRQs. The correlation between the broadband spectral indices obtained by Comastri et al. (1997) was interpreted to mean that there is a difference in shape of overall energy distributions of blazars from radio to X-ray energies. More detailed analyses of the broadband energy distributions of various blazar samples show that while most FSRQs and LSP-BL Lac have similar spectral properties, HSP-BL Lac appear to be different, which does not support a blazar sequence, but nevertheless, provide substantial evidence for a unified scheme for the blazars.

Although blazar unification through broadband spectral indices appears to have been well studied up to X-ray band by several authors (e.g. Nieppola et al. 2006; Fan et al. 2016), it is still poorly studied up to the *γ*-ray band due mainly to inadequacy of modelled *γ*-ray data for a vast majority of blazars. However, since *γ*-ray is almost exclusively due to inverse Compton scattering of relativistic particles in the jets it is becoming increasingly important in blazars for modelling their Compton spectra (Finke 2013; Dermer and Giebels 2016). In this paper, broadband spectral indices of a large sample of Fermi-LAT AGNs are statistically examined up to *γ*-ray band, with a view to understanding the nature of the relationship between synchrotron and Compton spectra of various subclasses of blazars.

## 2. Theory of Relationships

The low energy component of blazars, ranging from radio to X-ray is often attributed to synchrotron process while the highest energy end from X-ray to γ-ray is linked to inverse Compton process in the jets (e.g. Finke 2013). Thus, several authors adopted some proxy parameters, which depend on broad band spectral energy distribution, as well as frequency at synchrotron peak ($\nu_{pk}$), to study the blazar sequence (Fossati et al. 1998; Finke, 2013; Nalewajko and Gupta, 2017). Among the proxy parameters that are frequently used is the effective broad-band spectral index ($\alpha_{1-2}$), usually defined between two frequencies (e.g. Leddden and Odell 1985) as:

$$\alpha_{1-2} = -\frac{\log\left(\frac{F_1}{F_2}\right)}{\log\left(\frac{\nu_1}{\nu_2}\right)}, \qquad (1)$$

where $F_1$ and $F_2$ are the observed radiation fluxes at the frequencies $\nu_1$ and $\nu_2$ respectively. While there is considerable overlap of synchrotron and inverse Compton emissions in the X-rays for a large population of blazars, it is often convenient (e.g. Finke 2013) to refer to the radio-to-X-ray broadband spectral index ($\alpha_{rx}$) as synchrotron spectral index ($\alpha_{syn}$), because it arises mainly from the synchrotron spectrum. Similarly, X-ray to γ-ray spectral index ($\alpha_{x\gamma}$) is often (e.g. Finke, 2013) referred to as Compton spectral index ($\alpha_{com}$), because it arises mainly from inverse Compton spectrum. Although, it is now well known that extreme high energy peaked BL Lacs with synchrotron peaks up to GeV energy are emerging from recent observations, suggesting that GeV emissions are not pure Compton spectra, statistics of such objects are still limited (e.g. Foffano et al. 2019). In fact, only 55 objects in 3[rd] Fermi-Large Area Telescope (Fermi-LAT) blazar catalogue were reported to have been detected in TeV energy (Ackermann et al. 2015) and the TeV Compton spectra of these extreme objects are yet to be unambiguously modelled (Ackermann et al. 2015; Foffano et al. 2019). Hence, statistical population properties of γ-ray loud blazars still rely on the Fermi-LAT data, for which this assumed spectral modelling still remains valid.

The spectral luminosity ($P_\nu$) of a radio source at redshift ($z$) is related to its spectral flux density ($F_\nu$) at observing frequency ($\nu$) according to the relation (e.g. Ubachukwu *et al.* 1993)

$$P_\nu = F_\nu d_L^2 (1+z)^{\alpha_\nu - 1}, \tag{2}$$

where $d_L$ is the luminosity distance, which can be defined in the modern concordance cosmology in the form (e.g. Alhassan et al. 2019)

$$d_L = H_o^{-1} \int_0^z \left[(1+z)^2(1+\Omega_m z) - z(2+z)\Omega_\Lambda\right]^{-\frac{1}{2}} dz \tag{3}$$

where $\Omega_m$ and $\Omega_\Lambda$ are respectively, the contributions of baryonic matter and cosmological constant to the total energy of the universe ($\Omega_0 = \Omega_m + \Omega_\Lambda$).

However, for a complete sample with a flux density cut-off at $F_\nu = F_c$, equation (2) can be written in the form (e.g. Ubachukwu *et al.* 1993; Alhassan *et al.* 2013):

$$P_\nu = 4\pi d_L^2 F_\nu H(F_\nu - F_c)(1+z)^{\alpha_\nu - 1}, \tag{4}$$

where $H(F_\nu - F_c)$ is the Heaviside step function given by:

$$H(F_\nu - F_c) = \begin{cases} 0 & \text{if } F_\nu < F_c \\ 1 & \text{if } F_\nu > F_c \end{cases}. \tag{5}$$

Eq. (4) can be used to show a simple power law $P_\nu - z$ relation (e.g. Ubachukwu et al. 1993) as

$$\log P_\nu = \log P_o + \lambda \log(1+z) \tag{6}$$

where $P_0 \approx P_0(\Omega_0, H_o, c)$ and $\lambda \sim \alpha_\nu - 1$ is the slope of the $P_\nu - z$ data. Thus, in flux density limited samples, sources bunch up close to the flux limit due to Malmquist bias, with a strong correlation between $P_\nu$ and $z$.

Several authors (e.g. Alhassan *et al.* 2013; Odo *et al.* 2014) have shown theoretically and observationally that for low density universe, there is a steep change in $\lambda$ (and hence, $\alpha_\nu$) between low redshift ($z < 0.3$) and high redshift ($z \geq 0.3$) source samples. An obvious implication of all these is that the spectral index is redshift dependent (e.g. Athreya and Kapahi, 1998) and should be accounted for while using the parameter to study specific predictions of any unified scheme.

In the relativistic beaming model – a theoretical framework that has been remarkably successful in explaining the rapid variability of radiation flux from blazars, the observed spectral flux ($F_{\text{obs}}$) depends strongly on the viewing angle ($\phi$) and is related to the intrinsically emitted flux ($F_{\text{in}}$) by (Lind and Blandford 1985; Yuan-Tuan *et al.* 2016; Pei *et al.* 2019):

$$F_{\text{obs}} \approx \delta^{n+\alpha} F_{\text{in}}, \tag{7}$$

where *n* is a jet model dependent factor which is either 2 or 3 for continuous jet or blob models, respectively, and $\delta = [\Gamma(1-\beta\cos\phi)]^{-1}$ is the Doppler enhancement factor, with $\Gamma = (1-\beta^2)^{-\frac{1}{2}}$ being the Lorentz factor of jet material and *β* is the jet component speed in units of the speed of light.

Recent analyses (e.g. Chen *et al.* 2016; Odo *et al.* 2017) have provided evidence that orientation and relativistic beaming are imperative in explaining variations of high and low energy emissions from blazars. This is evidently in agreement with Giommi *et al.* (2013) that there is a relationship between high and low energy flux of blazar samples. However, the fact that high energy sources have small linear sizes implies that high energy emissions are from the core and are relativistically beamed (Savolainen *et al.* 2010). Therefore, we can represent using equation (7), a sequence of orientation of the jets to the line of sight using a systematic trend in variation of high energy flux from FSRQs at high energy flux to HSPs at low energy flux. In addition, the discovery of high energy neutrino flux associated with blazars (TXS 0506 + 056 for example) has released a scenario that will help comprehend the fundamental physical processes in blazars (Ice Cube Collaboration, 2018). The neutrino flux ($F_\mu$) is found to correlate significantly with high energy flux ($F_g$) in blazars which is expressed (e.g. Palladino *et al.* 2019) as:

$$F_\mu \approx \psi F_g \tag{9}$$

where ψ is a parameter that quantifies neutrino production efficiency from accelerated cosmic rays within the jet. The blazar radiation model (Keivan *et al.* 2018) and the neutrino emission around blazars can be fully explained by applying equation (7). A clear consequence of this scenario is that neutrino emission from blazars can also be accommodated in a beaming model of blazar unified scheme and should also give rise to a sequence across different subclasses of γ-ray loud blazars.

## 3. Data Sample and Results

From the third catalogue of Active Galactic Nuclei detected by the Fermi-Large Area Telescope (Fermi-LAT), compiled by Ackerman et al. (2015) and Acero et al. (2015), we select a sample of 1081 blazars with clear optical identification: 461 FSRQs and 620 BL Lacs. First, we collect the monochromatic luminosities of the objects from the available 1.4 GHz in radio, 1KeV in X-ray and I-GeV *Fermi*-LAT γ-ray data. In order to have a well-defined sample, we did not consider any candidate blazars of unknown type (BZU) from either sample. Most objects in current sample have been detected in various earlier surveys, which enabled a characterization of their spectral energy distributions using empirical relationships between low and high energy flux (Acero et al. 2015; Ackerman et al. 2015). There is considerable overlap in the data samples presented by the two groups of authors, and hence, for homogeneity, we take all relevant information, including SED classification and synchrotron peak frequency from Acero et al. (2015). Out of the 1081 blazars, only 680 objects have complete information on all three monochromatic luminosities of interest. The 680 objects which form the sample for current investigation include 279 FSRQs, 130 LSPs, 133 ISPs and 138 HSPs. For all objects in our sample, we calculate the *synchrotron and Compton spectral indices* from the available monochromatic properties. Throughout the paper, we adopt the modern concordance $(\Lambda - CDM)$ cosmology with $H_0 = 70$ kms$^{-1}$Mpc$^{-1}$ and $\Omega_0 = \Omega_m + \Omega_\Lambda = (\Omega_m = 0.3; \Omega_\Lambda = 0.7)$. The energy spectral index $\alpha_\nu$ is defined such that $F_\nu \sim \nu^{-\alpha}$. For a complete discussion of the unified scheme, we include eight (8) known young, jetted galaxies detected by Fermi-LAT, with available γ-ray luminosity and redshift data taken from Teng et al. (2011) and Rani (2019).

## 3.1 Distributions of γ-ray luminosity and spectral indices

Distributive analyses of the parameters of blazar samples are not only necessary in modelling their emission mechanisms but also essential in the study of the unification scheme. The distribution of γ-ray luminosity $(L_\gamma)$ is shown in Figure 1. Apparently, the distribution is continuous, with the FSRQs displaced to highest γ-ray luminosity and jetted galaxies to lowest γ-ray luminosity, suggestive that FSRQs are stronger γ-ray emitters than BL Lacs and radio galaxies. Nevertheless, the distributions yield mean (logarithm) values ~ 45.30 for FSRQs, 45.00 for LSPs, 44.50 for ISPs, 44.20 for HSPs and 42.88 for the galaxies, which follow the sequence $Log\langle L_\gamma\rangle|_{\text{FSRQs}} > Log\langle L_\gamma\rangle|_{\text{LSPs}} > Log\langle L_\gamma\rangle|_{\text{ISPs}} > Log\langle L_\gamma\rangle|_{\text{HSPs}} > Log\langle L_\gamma\rangle|_{\text{GALAXIES}}$ suggestive of a sequence that can extend to jetted galaxies. Statistical test reveals that FSRQs and BL Lac subclasses are nicely fitted to a log-normal distribution with skewness $(\mu)$ in the range $-0.08 \leq \mu \leq 0.04$. The observation suggests that similar mechanisms are responsible for the variations in γ-ray luminosity of all subclasses of blazars. Furthermore, two-sample Kolmogorov-Smirnoff (K-S) test carried out on the data shows that in general, at 95% confidence there is almost a zero probability ($\rho < 10^{-5}$) that there is any fundamental difference between the underlying distributions of these objects in γ-ray luminosity. Similarly, the jetted galaxies do not appear to be significantly different from FSRQs and BL Lac subclasses: there is only a probability $\rho \sim 0.05$ that the galaxies are different in terms of average luminosity distribution. The galaxies are observed to occupy the lowest luminosity regime of the distribution, which suggests that in general there is a connection between FSRQs, BL Lacs and the radio galaxies.

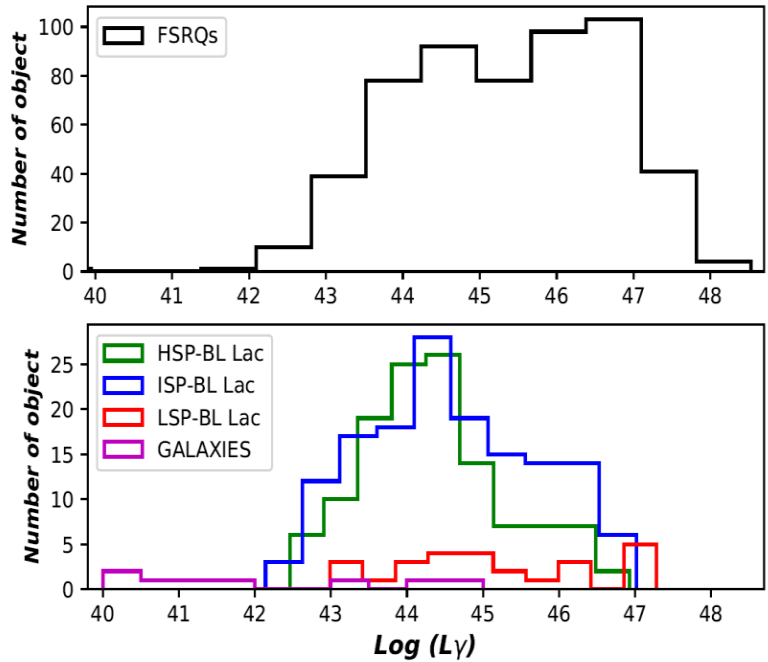

*Fig 1: Distributions of γ-ray luminosity FSRQs and BL Lac subclasses*

We show in Figure 2 the distributions of the sample in both Compton ($\alpha_{com}$) and synchrotron spectral index ($\alpha_{syn}$). It is however clear from Fig. 2(a) that BL Lacs, on average, have the tendency to possess higher values of Compton spectral index than the FSRQs. The average values are 0.60, 0.71, 0.78 and 1.10 for FSRQs, LSPs, ISPs and HSPs, respectively. Hence, the average values of the $\alpha_{com}$ for the blazar subclasses follow the relation: $\langle \alpha_{com} \rangle |_{HSPs} > \langle \alpha_{com} \rangle |_{ISPs} > \langle \alpha_{com} \rangle |_{LSPs} > \langle \alpha_{com} \rangle |_{FSRQs}$. However, FSRQs and BL Lacs are similarly distributed in multimodal configuration. Nevertheless, the multimodal distribution of BL Lacs arises from the fact that there are different subclasses of BL Lacs. For each subclass, it is essentially a normal distribution. This result can be interpreted to mean that in terms of Compton spectra, different subclasses of FSRQs may exist. In the case of $\alpha_{syn}$, FSRQs range from 0.78 to 1.30 with a single peak at 0.96 and average values of 0.98. However, BL Lacs range from 0.48 to 1.08 peaking at different values for LSP, ISP and HSP objects. The average values are 0.86, 0.84, and 0.66 for LSPs, ISPs and HSPs, respectively. We find that the average values of the $\alpha_{syn}$ for the blazar subclasses follow the relation: $\langle\alpha_{syn}\rangle|_{HSPs} < \langle\alpha_{syn}\rangle|_{ISPs} < \langle\alpha_{syn}\rangle|_{LSPs} < \langle\alpha_{syn}\rangle|_{FSRQs}$, a trend that is more or less a mirror image of the $\alpha_{com}$ distribution, which suggests a sequence for the blazar sample.

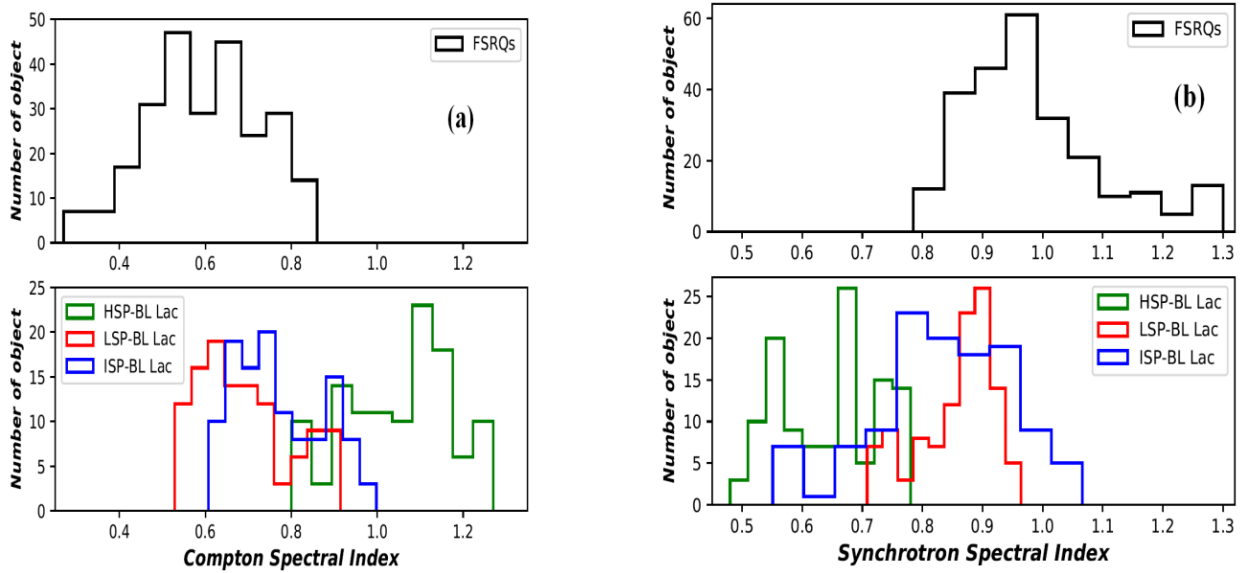

*Fig 2: Distributions of (a) Compton (b) Synchrotron spectral indices for FSRQs and BL Lacs*

We carried out *K–S* test on the spectral index data of the sample. The cumulative distribution functions (CDF) of the different subclasses are shown in Fig. 3. We find that for the Compton spectra, in Fig 3 (a), the overlap is less obvious, suggesting that the different subclasses of blazars may be statistically different in Compton spectral properties. Nevertheless, there is a sequence in the distribution of the Compton spectral index in a sense that is consistent with a blazar sequence. This can be understood in terms of a fundamental difference in the rate of cooling in these objects.

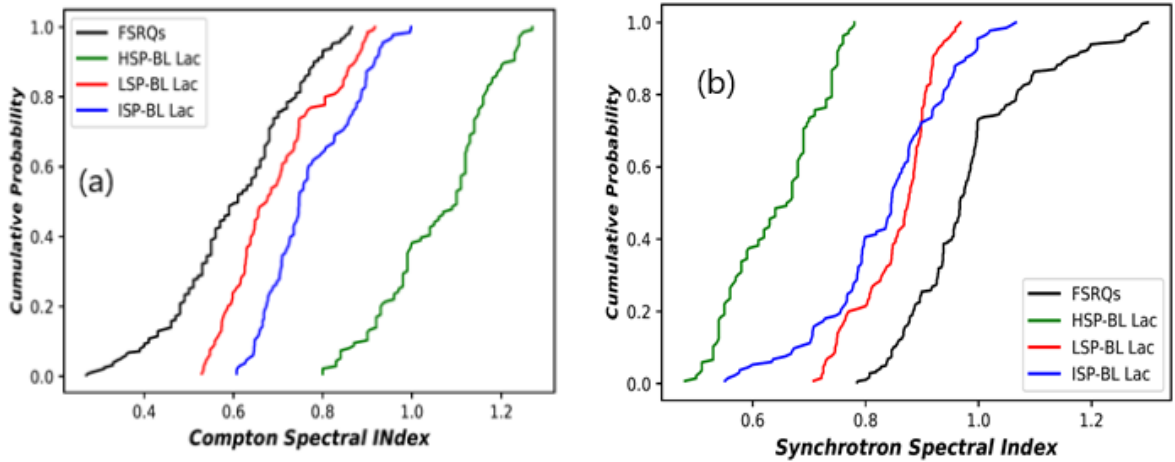

*Fig. 3: cumulative distributions of synchrotron (a) and Compton (b) spectral indices*

### 3.2 Correlation between parameters

To test for the correlations between the spectral parameters of the sample, the Pearson's correlation is applied. The $\alpha_{com} - \alpha_{syn}$ plot is shown in Figure 4. FSRQs are exclusively located at the low $\alpha_{com}$ high $\alpha_{syn}$ end while BL Lacs span a wide range, with LSP and ISP objects overlapping with FSRQs at intermediate values. The distribution of the objects on the $\alpha_{com} - \alpha_{syn}$ plane is in a sense that is consistent with blazar sequence with HSP-BL Lacs at high $\alpha_{com}$ - low $\alpha_{syn}$ through ISP and LSP to FSRQs at low $\alpha_{com}$ - high $\alpha_{syn}$ end. While most FSRQs have flat Compton spectral $|\alpha_{com}| > 0.5$, BL Lacs show steep Compton spectra $|\alpha_{com}| < 0.5$ which has a wide range. This is an indication of a range of inverse Compton scattering mechanisms in BL Lacs. It can be argued that the recurrent tendency observed from FSRQs up to HSP-BL Lacs follows a system which is indicative of the orientation unified scheme and that related mechanisms control these objects. Linear regression analysis of our data for the whole sample taken together gives $\alpha_{com} = -(1.08 \pm 0.02)\alpha_{syn} + (1.27 \pm 0.30)$ with a correlation coefficient ($r$ = -0.84) at 95% confidence. The strong correlation suggests that similar effects are responsible for variations in the parameters for both BL Lacs and FSRQs but depends on a factor that changes progressively from HSPs to FSRQs through ISPs and LSPs.

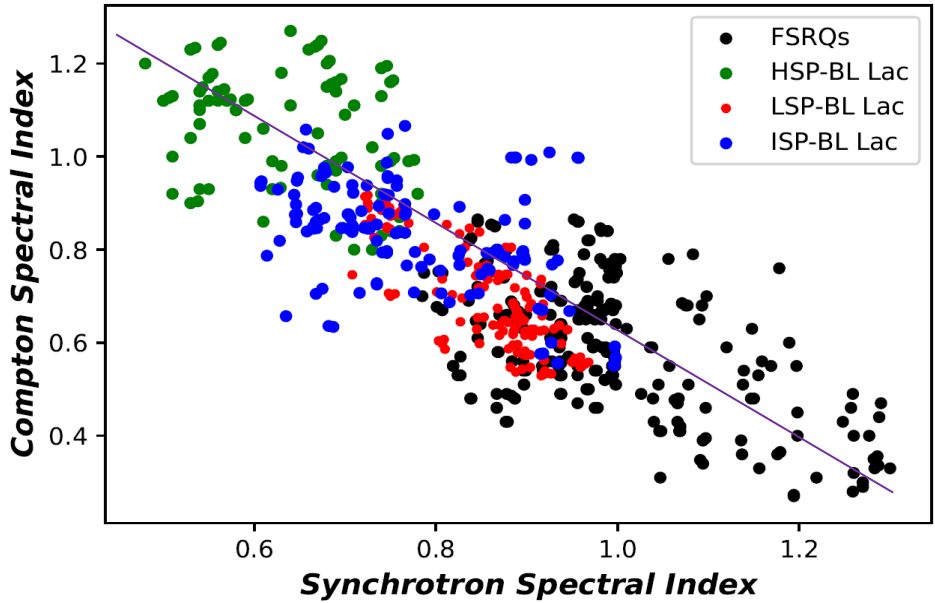

*Fig. 3: scatter plot of $\alpha_{syn} - \alpha_{com}$ for FSRQs and BL Lacs subclasses*

To investigate the relationships between the broadband spectral indices and synchrotron peak frequency, the scatter plots of $\alpha_{com}$-$\nu_{peak}$ and $\alpha_{syn}$-$\nu_{peak}$ are shown in Fig. 5. The FSRQs-LSP-ISP-HSP sequence is obvious in each plot. Apparently, there is a positive correlation of $\alpha_{com} - \nu_{peak}$ data for the entire blazar sample, which upturns into an anti-correlation on $\alpha_{syn} - \nu_{peak}$ plane. This shows that on $\alpha - \nu_{peak}$ plane, these objects follow a trend which is suggestive of a unified scheme through blazar sequence. Linear regression analyses of our data yield $\alpha_{com} = (0.27 \pm 0.03)\,\nu_{peak} + (0.37 \pm 0.20)$ with correlation coefficient $r$ = 0.84 and $\alpha_{syn} = -(0.22 \pm 0.03)\,\nu_{peak} + (1.28 \pm 0.30)$ with $r$ = -0.81 at 95% confidence. Observe that the correlations exist only when the blazars are taken together and becomes less important when individual subsamples are considered.

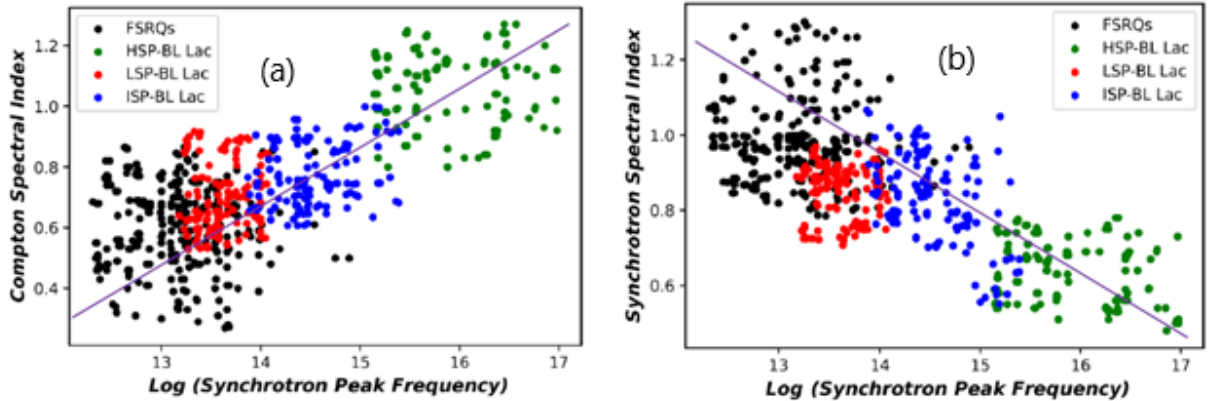

Fig. 5: Scatter plot of (a) $\alpha_{com}$ (b) $\alpha_{syn}$ against $\nu_{peak}$ for FSRQs and BL Lac subclasses

To investigate the effects of redshift on our results, we show the $\alpha_{com} - z$ and $\alpha_{syn} - z$ plots for the sample in Fig. 6. It is observed that there is a steep change in $\alpha - z$ around $z = 0.3$ in each case, shown with broken lines in Fig. 6. While all BL Lacs are located at low (z < 0.3) redshift, a vast majority of FSRQs are located at high (z > 0.3) redshift. Nevertheless, there is a smooth transition from HSP- BL Lac at lowest redshifts to FSRQs at high redshift end through ISP and LSP- BL Lacs that are almost indistinguishable on the $\alpha - z$ plane. This trend suggests that there is a unified scheme for blazars through blazar sequence or/and evolution.

Regression analysis of the $\alpha_{com} - z$ data yields $\alpha_{com} = (-0.26 \pm 0.02)\log(1+z) + (1.23 \pm 0.10)$ with (r = -0.65) for $z < 0.3$ and $\alpha_{com} = (-0.08 \pm 0.02)\log(1+z) + (0.81 \pm 0.20)$ with (r = -0.44) for $z \geq 0.30$. This implies a steep change in slope from -0.26 for $z < 0.3$ to -0.08 for $z \geq 0.3$. Similarly, for $\alpha_{syn} - z$ data, the result yields $\alpha_{syn} = (0.23 \pm 0.02)\log(1+z) + (0.47 \pm 0.20)$ with (r = 0.61) for $z < 0.30$ and $\alpha_{syn} = (0.09 \pm 0.01)\log(1+z) + (0.84 \pm 0.10)$ with (r = 0.47) for $z \geq 0.30$. There is also an indication of a change of slope from 0.23 at $z < 0.30$ to 0.09 at $z \geq 0.30$. The results indicate that the broadband spectral indices are fairly constant at $z \geq 0.30$

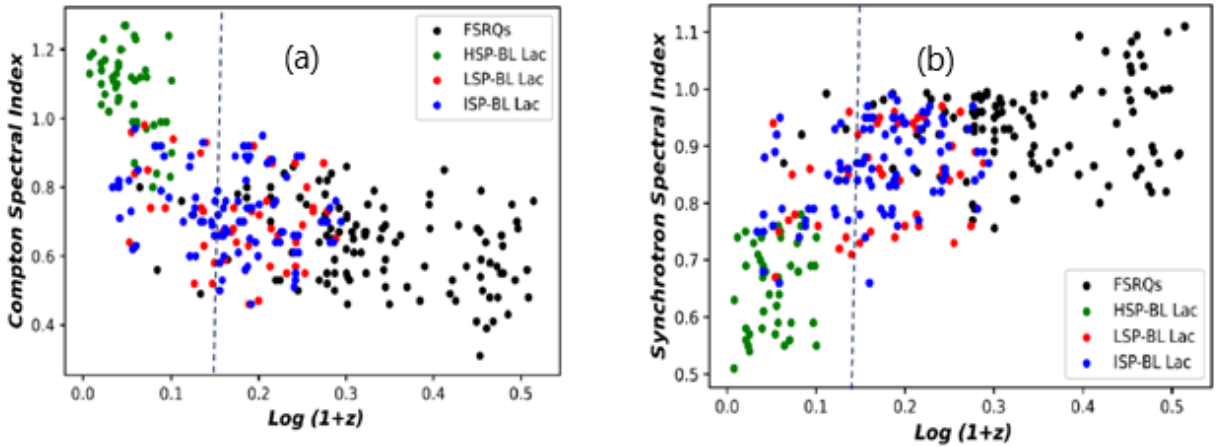

*Fig. 6: scatter plots of (a) $\alpha_{com}$ (b) $\alpha_{syn}$ as a function of log (1+z) for FSRQs and BL Las subclasses*

## 4. Discussion

The results of the current investigation of blazar sequence using homogenous sample of Fermi-LAT blazars are presented. The well-sampled data has allowed an investigation of the relationship between FSRQs and BL Lac subclasses. The theoretical blazar sequence is remarkably popular in that it unites distinct subclasses of blazars into one family (Fossati *et al.* 1998). This blazar sequence is seen in the light of relativistic beaming model which proposes a decrease in the Doppler boosting as the synchrotron peak frequency increases due to a drop in the luminosity (see e.g Meyer *et al.* 2011). Therefore, for any difference in FSRQs and BL Lacs due to orientation effect, the spectral energy distributions have to be considered in the beaming model. However, the systematics of the spectral energy distributions and blazar sequence originally proposed by Fossati *et al.* (1998) based on the 126 samples of the slew survey and 1Jy samples discovered that the different kinds of blazars follow a continuous spectral sequence. The theoretical blazar sequence also visited by Ghisellini and Tavachio (2008) revealed that the spectral energy distributions of blazar emissions as well as the power of the jet are all linked to the key parameters of the accretion process. The results have shown that both BL Lacs and FSRQs are sites of high energy phenomena and this supports the hypothesis of highly relativistic jets at small orientation angles to the line of sight (e.g. Mukherjee *et al.* 1997). The blazar sequence postulates that FSRQs and BL Lacs are different expressions of the same physical process that vary only by bolometric luminosity (Ghisellini *et al.* 1998: Fossati *et al.* 1998). Consequently, there should be continuity in distributions of spectral properties of these objects.

The distributions of our sample in γ-ray luminosity show that FSRQs are on average more prominent than BL Lacs. However, the strong γ-ray luminosity of FSRQs is indicative of the strong beaming effect which denotes that FSRQs are sources with highly boosted γ-ray continuum (e.g. Padovani 1992). Perhaps, the differences in the distributions of γ-ray luminosity of the blazar subclasses could be interpreted to mean that the two blazar subclasses differ in orientation (Yuan-Tuan *et al.* 2016; Chen *et al.* 2016).

Fig. 1 has an interesting feature in that the distribution of BL Lacs and FSRQs shows a continuous sequence in γ-ray luminosity, with no clear dichotomy between the various subclasses, which is suggestive that inherently similar processes control γ-ray emission in these objects. The distributions of our sample in broadband spectral indices also follow a continuous trend from FSRQs to HSPs through LSPs and ISPs in agreement with the blazar sequence. The variations in the spectral indices between FSRQs and BL Lacs can be understood in the framework of the unified scheme if a common factor which changes progressively between the subclasses controls the variation. The continuous distributions shown in Figure 2 thus indicates that there is a unification scheme for the broadband spectral properties of blazars, which is in close agreement with blazar sequence as proposed by Ghisellini *et al* .(1998) and Fossati *et al.*(1998). We interpret this to mean that the mechanisms producing the emissions in the two populations are similar but vary systematically between the subclasses (e.g. Chen *et al.* 2016).

On $\alpha_{com}$ - $\alpha_{syn}$ plane, it is shown that FSRQs and HSP-BL Lacs form the opposite extremes of the continuum connected by LSP-BL Lac and ISP-BL Lac populations that are indistinguishable. In fact, there is a considerable overlap between FSRQ and LBL populations on the $\alpha_{com}$ - $\alpha_{syn}$ plane. FSRQs have larger values of synchrotron spectral index but smaller values of Compton spectral index than the BL Lac subclasses. FSRQs, LSP-BL Lac and ISP-BL Lac have relatively low Compton spectral index compared to HSPs. This result is quite comparable with the anti-correlation of broadband radio-to-optical ($α_{ro}$) and optical-to -X-ray ($α_{ox}$) spectral indices (e.g. Fan et al. 2016; Ackermann et al. 2011; Abdo et al. 2010a) in which FSRQs and LSPs possess similar spectral properties that are significantly different from those of HSPs, but with ISPs forming a bridge connecting them. Actually, Comastri et al. (1997) discovered that there is a significant anti-correlation between X-ray and γ-ray spectral indices, and also between $α_{ro}$ and $α_{ox}$ broadband spectral indices of BL Lacs and FSRQs of SDSS blazars, which the authors explained in terms of a difference in shape of overall energy distributions from radio to X-ray energies. If this is actually the case, the $\alpha_{com}$ - $\alpha_{syn}$ anti-correlation observed in current investigation can be interpreted to mean that there is a difference in the shape of overall synchrotron energy distributions from those of Compton energies for all blazars. Interestingly, both $\alpha_{com} - v_{peak}$ and $\alpha_{syn} - v_{peak}$ correlation/anti-correlation observed in current analyses are consistent with a unification of the blazars through blazar sequence.

It is arguable that the $\alpha_{com}$ - $\alpha_{syn}$ anti-correlation observed in current investigation may have risen from a redshift dependence of both $\alpha_{com}$ and $\alpha_{syn}$. The results of our analyses show that the slopes of $\alpha_{com} - z$ and $\alpha_{syn} - z$ change from -0.26 to -0.08 and 0.23 to 0.09 at low-and high redshifts, respectively. It is interesting to observe that the steep change in α – z slope did not yield any noticeable effect on the $\alpha_{com}$ - $\alpha_{syn}$ anti-correlation. Thus, the $\alpha_{com}$ - $\alpha_{syn}$ anti-correlation is intrinsic and not an artefact of a common dependence of both parameters on redshift. Perhaps, the steep α – z slope for z < 0.3 objects could mean that at low redshifts, luminosity selection effect can be playing a significant role (e.g. Odo *et al.* 2014; Alhassan *et al.* 2013). Actually, Alhassan *et al.* (2013) opined that at low redshifts ($z \geq 0.30$), the effects of luminosity selection are so strong as to swamp any effect that could arise from relativistic beaming in a sample of BL Lacs and radio galaxies. Obviously, all the BL Lacs in the current sample are located at low redshifts (z < 0.3). If this is actually the case, the variations of synchrotron and Compton spectral indices at low redshifts may be an artefact of the luminosity selection effect. Furthermore, the steep change in α – z slope for $z < 0.3$ and $z \geq 0.3$ corresponding to regions occupied by BL Lacs and FSRQs, respectively, could be interpreted to mean that BL Lacs and FSRQs have distinct spectral properties. Interestingly, the presence of some FSRQs whose properties well overlap with those of BL Lacs on the α – z plane in Fig. 6 can be used to argue against any spectral dichotomy between the two populations of blazars and suggests that for all blazar classes, similar processes run and there is a unified scheme (e.g. Odo and Aroh, 2020). Li *et al.* (2015) earlier argued that the $\alpha_{ro} - z$ correlation using SDSS data provides more evidence for the unified scheme as proposed by Fossati *et al.* (1998).

It is also observed from fig 5 that the broadband spectral indices remain fairly constant at $z \geq 0.30$. The significant correlation in the broadband spectral indices against synchrotron peak frequency of the blazar subclasses suggests that similar processes run in all these objects within central physical environments. Thus, differences between blazar subclasses observed in earlier investigations (Comastri et al. 1997) can be attributed to intrinsically different rates of cooling at different positions within the cores of the various subclasses of blazars (e.g. Ghisellini et al. 1998; Savolainen *et al.* 2010). The synchrotron spectral index peaks at low and high synchrotron peak frequencies for FSRQs and HSP-BL Lacs, respectively, implying very strong and weak cooling levels for these objects (e.g. Ghisellini et al. 1998). Thus, if the synchrotron peak frequency is low for FSRQs, it would be expected that the Compton spectral index would equally be low for FSRQs and high for BL Lacs, which is consistent with our results.

We have noted the presence of few FSRQs (10% of the FSRQs in the sample) that share similar parameter space in luminosity and redshift with BL Lacs. This minority of FSRQs occupy the parameter space in such a manner that is consistent with a unified scheme between BL Lacs and FSRQs (e.g. Odo and Aroh, 2020). Apparently, the transition from HSP-BL Lacs at low z and high $\alpha_{com}$ to the FSRQs at high z and low $\alpha_{com}$ is consistent with an evolution-based unified scheme for BL Lacs and FSRQs in which at inverse Compton energies, blazars start off as HSP BL Lacs and evolve through ISPs and LSPs to FSRQs at high redshifts (Beckmann et al. 2003). The presence of the minority of FSRQs that share the same parameter space with BL Lacs in this analysis does not alter the evolutionary trend from HSPs to FSRQs through ISPs and LSPs which can be used to argue for a unified scheme for BL Lacs and FSRQs through evolution (e.g. Ajello *et al.* 2014). Perhaps, these few FSRQs could be candidate masquerading BL Lacs, whose presence are becoming increasingly important in recent investigations (e.g. Giommi et al. 2013; Padovani et al. 2019; Odo et al. 2020). It can therefore be argued that these BL Lac-like FSRQs could be a distinct subclass of FSRQs.

Another important outcome of our analyses is the position of jetted radio galaxies in the current unification scheme in γ-ray band. it is important to note that the location of the jetted galaxies in redshift and γ-ray luminosity is also consistent with both evolutionary and orientation sequence as they are located at much lower redshift ($z \sim 0.006$) than HSP-BL Lacs and form the low luminosity tail in the γ-ray distribution in Fig.1. Actually, Rani (2019) argued that the radio morphology of the jetted galaxies in their sample resemble blazars when viewed at larger angles to the line of sight. The observed unification sequence presented in this paper, thus, is expected to include these misaligned jetted γ-ray emitting radio galaxies. However, statistics are still poor for these radio galaxies in the γ-ray band.

## 5. Conclusion

We have used the distributions and correlations of the observed parameters of extragalactic radio sources to quantitatively test a simple statistical consequence of the spectral properties of some *Fermi* selected blazars. We showed that $\alpha_{com}$ and $\alpha_{syn}$ of the two blazar subsets (FSRQs and BL Lacs) are strongly anti-correlated ($r \sim -0.80$) for our sample, which we interpret as significant difference in shape of synchrotron and Compton spectra. Distribution of Compton spectral indices suggests that different subclasses of FSRQs can be possible. The distributions of the observed parameters of our sample suggest a smooth transition from FSRQ to HSPs through LSPs and ISPs. Our results thus strongly support the earlier suppositions for a unified scheme for blazars. We also found that the blazar sequence can be extended to the young jetted radio galaxy populations whose γ-ray emissions are gaining attention in recent investigations.

**Acknowledgement**
We sincerely thank an anonymous referee whose invaluable comments and suggestions helped to improve the manuscript. This work is financially supported by the National Economic Empowerment and Development Strategy of the Federal Government of Nigeria (NEEDs) and partly done at the North-West University, Potchefstroom, South Africa.

**REFERENCES**
Abdo, A. A., Ackermann, M., Ajello, M., Atwood, W. B. *et al.* 2009, *ApJ*. 700, 597
Abdo, A. A., Ackermann, M., Agudo, I., Ajello, M. *et al.* 2010c, *ApJ*. 716, 30
Abdo, A. A., Ackermann, M., Ajello, M., Atwood, W.B. et al. 2010a, *ApJ*. 710, 1271
Abdo, A. A., Ackermann, M., Ajello, M., Allafort, A. et al 2010b, *ApJ*. 715, 429
Ackermann, M., Ajello, M., Atwood, W.B., Baldri, E. et al. 2015.*ApJ*. 810: 14
Ackermann, M., Ajello, M., Allafort, A. et al. 2011.*ApJ*. 743: 171
Aero, F, Ackermann, M., Ajello, M. et al. 2015, *Astrophys. J. Sup*, 218, 23
Ajello, M., Romani, R. W., Gasparrini, D. et al. 2014. *ApJ*. 780,73
Alhassan, J.A., Ubachukwu, A.A., Odo, F.C. and Onuchukwu, C. C. 2019, *Rev MexAA*. 55, 151
Alhassan, J.A., Ubachukwu, A.A., Odo, F.C. 2013. *Jour. of Astrophys & Astron,* 34, 61
Andruchow, I., Romero, G.E., Cellone, S.A., 2005. *Jour. of Astron .& Astrophys.*, 442, 97
Athreya, R. M., Kapahi, V. K. 1998.*Jour. of Astrophys. Astron,* 19, 63
Blandford, R. D., Rees, M. J. 1978. In Pittsburgh Conference on BL Lac Objects, ed. A. M. Wolfe (Pittsburgh: Univ. Pittsburgh Press), 328

Beckmann, V., Engles, D., Bade, N., Wucknitz, O. 2003, *Astron. Astrophys.*, 401, 927.
Böttcher, M. 2007. *Astrophys. & Space Sci.*, 309, 95
Cellone, S.A., Romero, G.E., Araudo. A. T., 2007, *Mon. Not. R. Astron. Soc.* 374, 357-364
Chen, Y. Y., Zhang, X., Xiong, D. R., Wang, S. J.,Yu, X. L. 2016. *Res. Astron. Astrophys.* 16, 13
Comastri, A., Fossati, G., Ghisellini, G., Molendi, S. 1997. *ApJ.* 480, 534
Dermer, C. D., Giebels, B. 2016, C. R. Physique, 17 594.
Fan, J. H., Yang. J.H., Yuan, Y.H., Wang, J., Gao, Y., 2012. *ApJ.* 761:125
Fan, J. H., Yang, J. H., Liu, Y., et al. 2016, *ApJS*, 226, 20
Finke, J. D. 2013. *ApJ*, 763, 134
Foffano,L., Prandini, E., Franceschini, A. and Paiano, S. 2019, arXiv:1903.07972v2 [astro-ph.HE]
Fossati, G., Maraschi, L., Celotti, A., Comastri, A., Ghisellini, G. 1998. *Mon. Not. R. Astron. Soc*. 299:433
Gaur, H., Gupta, A.C., Lachowerez, P., Wiita, P. 2010. *ApJ.* 718, 279
Ghisellini, G., Celotti, A., Fossati, G., Maraschi, L., Comastri, A. 1998.*Mont. Notic. of the Royal Soc.* 301, 451
Ghisellini G., Tavecchio F., 2008, *Mont. Notice of the R. Soc*. 387, 1669
Giommi, P., Padovani, P., Polenta, G. 2013. *Mon. Not. R. Astron. Soc*. 431, 1914
Giommi, P., Padovani, P., Polenta, G., Turriziani, S., D'Elia, V. et al. 2012, *Mon. Not R. Astron. Soc.* 420, 2899
Giommi, P., Ansari, S. G., & Micol, A. 1995, *A&AS,* 109, 267
IceCube Collaboration, 2018. *Science,* 361, 147
Keivan, A., Murase, K., Petropoulou, M., Fox, D.B. 2018. *ApJ*. 864, 84
Kharb, P., Lister, M. L., Cooper, N. J. 2010. *ApJ*. 710, 764
Kollgaard, R.I. 1994. *Vistas in Astronomy,* 38, 29
Ledden, J.E., Odell, S.L. 1985. *ApJ.* 298, 630-643
Li, H. Z., Xie, G. Z., Yi, T. F., Chen, L. E. *et al.* 2010. *ApJ.* 709, 1407
Li, H. Z., Cheng, L.E., Jiang, T.F. 2015. *Res. Astron.& Astrophys.* 15(7), 929
Lind K. R., Blandford R. D.1985. *ApJ.* 295, 358
Lister, M.L. Aller, M., Aller, H. Hovatta, T. *et al.* 2011, *ApJ.* 742, 27
Meyer, E. T., Fossati G., Georganopoulos, M., Lister M. L. 2011. *ApJ,* 740, 98
Mukherjee, R., Bertsch, D. L, Bloom, S. D., Dingus, B. L. et al. 1997, *ApJ.* 490, 116
Nalewajko, K., & Gupta, M. 2017. *Astro & Astro*, 606, A44
Nieppola, E., Tornikoski, M., & Valtaoja, E. 2006. *Astron. Astrophys.*, **445**, 441
Odo, F. C. and Aroh, B. E. 2020, *J. Astrophys. Astron.* 41, 9
Odo F. C., Chukwude, A. E., Ubachukwu, A. A. 2017, *Astrophys & Space Sci.*, 362, 23
Odo F.C., Chukwude A.E., Ubachukwu A.A. 2014.*Astrophys & Space Sci.*, 349, 939
Padovani, P., Oikonomou, F, Petropoulou, M., Giommi, P., Resconi, E. *MNRAS* Letters, 484, 5 pp. (arXiv:1901.06998) (2019)
Padovani, P. 2007. *Astrophys & Space Sci.* 309, 63
Padovani, P. 1992. *Mont. Notice of the R. Soc*. 257, 404

Palladino, A., Rodrigues, X., Gao, S., Winter, W. 2019. *ApJ.* 871; 41

Pei Z. Y., Fan, J. H., Bastieri, D., Sawangwit, U., Yang, J H. 2019, *Res. Astron. Astrophys.* 19, 70

Raiteri, C. M., Capetti, A., 2016. *Jour. of Astron. & Astrophys.* 587, A8

Rani, B., 2019, Galaxies, 7, 23

Sambruna, R. M., Maraschi, L., Urry, C. M. 1996. *ApJ.* 463, 444

Savolainen, T., Homan, D.C., Hovatta, T., Kadler, M. et al. 2010. *Jour. of Astro & Astro.* 512, A24

Scarpa R. & Falomo R. 1997. *Jour. of Astro. & Astro.*, 325, 109

Teng, S.H., Mushotxky, R.F., Sambruna, R.M., Davies, D.S. et al. 2011. *ApJ.* 742, 66

Ubachukwu, A.A., Okoye, S.E. & Onuora, L.I. 1993. *Proceedings of Nig. Academy of Sci.* 5, 63

Urry, C. M. 2011, *Jour. of Astrophys. & Astron.,* 32, 139-145

Urry, C.M., Padovani, P., 1995. *PASP*, 107, 803

Xie, G.-Z., Dai, B.-Z., Mei, D.-C., Fan, J.-H. 2001, *Chin. J. Astron. Astrophys.*, 1, 213-220

Yang, J., Fan, J., Liu, Y., et al., 2018. *China Physics, Mechanics and Astronomy,* 61, 059511

Yuan-Tuan Li, Fu, S.Y., Feng, H.J., He. S.L., et al. 2016, *Jour. of Astrophys & Astron.* 123.

Yefimov, Yu.S. 2011. *Jour. of Astrophys. & Astron.*, 32, 73